\documentstyle[12pt]{article}
\def\beq{\begin{equation}}
\def\eeq{\end{equation}}
\def\bea{\begin{eqnarray}}
\def\eea{\end{eqnarray}}
\def\bq{\begin{quote}}
\def\eq{\end{quote}}

\def\gappeq{\mathrel{\rlap {\raise.5ex\hbox{$>$}}
{\lower.5ex\hbox{$\sim$}}}}

\def\lappeq{\mathrel{\rlap{\raise.5ex\hbox{$<$}}
{\lower.5ex\hbox{$\sim$}}}}
\begin{document}
\topmargin -0.5cm
\oddsidemargin -0.8cm
\evensidemargin -0.8cm
\pagestyle{empty}
 \begin{flushright}
CERN-TH.7224/94\\
IAS 94/24
\end{flushright}
\vspace*{5mm}
\begin{center}
{\bf GAUGE FIELD IMPROVEMENT, FORM-SCALAR DUALITY,}\\
{\bf CONFORMAL INVARIANCE AND QUASI-LOCALITY}\\
\vspace*{1cm}
 S. Deser\footnote{On leave from Physics Dept., Brandeis Univ.,
Waltham, MA 02254, U.S.A.}\\ Theoretical Physics Division, CERN, Geneva,
Switzerland\\ \vspace {0.2cm}
 and\\
School of Natural Sciences, Institute for Advanced Study\\
Princeton, NJ 08540, U.S.A.\\
\vspace {0.4cm}
and\\
\vspace {0.4cm}
 A. Schwimmer\\ Weizmann Institute, Rehovot, Israel and SISSA and
  INFN, Trieste,Italy \\
 \vspace{2.0cm}
Abstract
\end{center}
The problem of maintaining scale and conformal invariance in Maxwell and
general
$N$-form gauge theories away from their critical dimension $d_c =
2(N+1)$ is
analyzed.  We first exhibit the underlying group-theoretical clash
between
locality, gauge, Lorentz and conformal invariance requirements.
``Improved" - traceless - stress tensors are then constructed;  each
violates one
of the above criteria.  However, when $d = N+2$, there is a duality
equivalence
between $N$-form models and massless scalars .              Here we show
that $N$-form
conformal invariance is not lost, by constructing a quasilocal gauge
invariant
improved stress tensor.  The correlators of the scalar theory are then
reproduced, including the latter's trace anomaly.  This resolves the old
conformal invariance ``paradox" between the dual partners.
\vspace*{2.0cm}
\begin{flushleft}
CERN-TH.7224/94\\
April 1994
\end{flushleft}
\vfill\eject
\pagestyle{empty}
\clearpage\mbox{}\clearpage
\setcounter{page}{1}
\pagestyle{plain}

\section{Introduction}
Massless spin-1/2 fields have vanishing classical stress tensor trace in
any
dimension, while scalars can be ``improved" to achieve $T^\alpha_\alpha
= 0$,
thereby guaranteeing invariance under the special conformal (or full
Weyl) group,
in accord with their scale-independence.  This is no longer the case for
the
Maxwell and general $N$-form gauge fields' stress tensor traces:  they
only vanish
at critical dimension $d_c = 2(N+1)$.  It is therefore instructive to
analyse the
obstacles to improvement away from $d_c$.  We shall first give a simple
general
group theoretical argument exhibiting the clash between gauge and
conformal
transformations.  This argument involves the sensible assumptions of
locality and
Lorentz invariance.  We will then construct explicit traceless stress
tensors away
from $d_c$, and see that they indeed violate gauge or Lorentz
invariance.
[Non-local improvement is trivial to accomplish, of course.]

A particularly interesting dimension is $d=N+2$ where there is a duality
equivalence between $N$-forms and scalars.  Here there is a longstanding
apparent
paradox:  the scalar's stress tensor can of course be improved, whereas
we have
shown that the $N$-form's cannot.  The resolution lies in
the presence of a quasilocal (local on-shell) improvement
for the latter that leads to
equivalence of correlation functions in both languages, including in
particular
equality of trace anomalies of these - equally conformal invariant -
dual models.

Most of our discussion will take place in the framework of free theories
in flat
space, the issue being preservation of the full conformal group.
Equivalently,
one could work in an external gravitational field where this translates
to
preservation of the Weyl group.  We shall not be concerned here with
delicate
issues of zero-modes, addressing the main question of topologically
trivial spaces.

\section{Group Theory}
A massless Lorentz invariant field theory is invariant under scale
transformations
\beq
x^{\mu'} = \lambda x^\mu~,
\label{1}
\eeq
the basic field of the theory $\varphi_j(x)$ transforming as
\beq
\varphi_j' (x') = \lambda^{-\eta} {\varphi_j}(x)
\label{2}
\eeq
where $\eta$ is the dimension of the field, whose Lorentz properties are
encoded in the index $j$.
Correspondingly, there is a conserved dilatation current $D^\mu$ and
charge $D =
\int d^{d-1}xD^0$.  If $D^\mu$ can be
written in terms of a symmetric conserved (stress) tensor as
\beq
D^\mu = x^\nu T^{\mu}_{\nu}~,
\label{3}
\eeq
then conservation of $D^\mu$ implies
\beq
T^\mu_\mu = 0~.
\label{4}
\eeq
The conformal currents $K^{\mu(\alpha)}$,
\beq
K^{\mu(\alpha)} = (x^2g^{\alpha\nu} - 2x^\nu x^\alpha)T^\mu_\nu
\label{5}
\eeq
are then also conserved;  the charges $K^\alpha$ generate the special
conformal
transformations
\beq
x^{\mu'} = \frac{x^\mu + b^\mu x^2}{1 + 2bx + b^2x^2}
\label{6}
\eeq
on $\varphi_j$, corresponding to a symmetry of the theory.  Now, since
the special
conformal transformations are generated by translations and
the inversion
\beq
x'_\mu = \frac{x_\mu}{x^2}~,
\label{7}
\eeq
a necessary condition for the existence of a traceless energy momentum
tensor is
the invariance of the Lagrangian under (\ref{7}).  The transformation
of the
field under inversion
 is fixed by the conformal group to be
\beq
\varphi'_j(x') = (x^2)^\eta D^k_j \left(g^{\mu\nu} - 2 x^{-2}x^\mu
x^\nu \right) \varphi_k(x)
\label{8}
\eeq
where $D^k_j$ are the SO(d)-Wigner functions in the appropriate
representation evaluated for the SO(d) rotation in the argument of $D$
and $\eta$ is
the scale dimension of (2).

Massless fields with vectorial or higher rank indices are described
covariantly by
actions which possess a gauge invariance.  We limit ourselves here to
systems
described by bosonic $N$-forms $\varphi_{\mu_1...\mu_N}$. [For   recent
general reviews of conformal invariance see \cite{aaa} \cite{dd}.]
The gauge field
action
\beq S = \int *d(\varphi *) d\varphi
\label{9}
\eeq
is invariant under gauge  transformations defined by
the $N-1$ form $\alpha$,
\beq
\varphi \rightarrow \varphi + d \alpha~.
\label{10}
\eeq
It is also invariant under dilations if we choose
\beq
\eta = \frac{1}{2} (d-2)~,
\label{11}
\eeq
while dilations commute with the gauge transformation provided
\beq
\eta_\alpha = \eta_\varphi - 1.
\label{12}
\eeq
As argued above, in order to have a traceless energy momentum tensor,
the
action (\ref{9}) should be invariant under the transformation
(\ref{7}), (\ref{8}) with $\eta$ given by (\ref{11}).  For if the action
is
invariant under the inversion (\ref{7}) and the inversion does not
commute with the
gauge transformations,then (\ref{9}) will be invariant also under a new
gauge group
generated by inversion  times gauge transformation times inversion. So
the
gauge transformations (\ref{10}) must commute with conformal
transformations
(\ref{7}), (\ref{8}). Now, the effect of diffeomorphisms $x^\mu
\rightarrow
x^{\mu'}(x)$ is
$$
\varphi(x)dx \wedge ... \wedge dx = \varphi'(x')dx'\wedge ... \wedge dx'
\eqno{(13a)}
$$
$$
\alpha(x)dx \wedge  ... \wedge dx = \alpha'(x') dx' \wedge ...dx'~,
\eqno{(13b})
$$
which automatically commute with (\ref{10}).  The conformal group
considered as a
subgroup of diffeomorphisms will therefore commute with (\ref{10}),
but then the dimension $\eta$ is
clearly fixed by (13) to be
\addtocounter{equation}{1}
\beq
\quad \eta_\varphi = N~, \quad \eta_\alpha = N-1~.
\label{14}
\eeq
  This requirement is equivalent to the form being inert under
a Weyl transformation.
Therefore the action for an $N$-form in $d$ dimensions will be invariant
under
the full conformal group
momentum tensor iff
\beq
d = d_c = 2(N+1)~.
\label{15}
\eeq
At $d_c$, the conformal charges will be gauge invariant automatically.
The above
arguments assumed implicitly that the gauge fields transform in the
usual, local
way under the conformal group.  This assumption will be important in our
analysis
of $N$-form-scalar duality in Section~4, to explain the apparent
discrepancy in
their conformal behaviour at $d = N+2$.

In light of the above general result, there can be no strictly local
(but see
Section~IV), Lorentz and gauge invariant conserved traceless stress
tensor away
from $d_c$.  Hence we have two options if we insist on improvement:
\begin{itemize}
\item[a)] Work entirely in terms of the non-covariant but
gauge-invariant
variables.  [Although gauge-fixing is not strictly necessary in this
Abelian
framework, one may think in terms of a physical ``Coulomb" gauge choice
here.]
This will, however, entail loss of Lorentz invariance.
\item[b)] Choose a covariant (Landau or Feyman) gauge;  the resulting
generators
will depend on unphysical (ghost) degrees of freedom, due to loss of
gauge
invariance.
\end{itemize}
\section{Explicit Improvement}

We now turn to explicit realizations of condition (\ref{4}),
$T^\alpha_\alpha =0$ for our systems.  Recall that the possibility of
improving a
stress tensor rests on its non-uniqueness since it is not a physical
local
current in the absence of gravitational coupling;  one may add to
$T^{\mu\nu}$
any quantity of the form:
\beq \Delta^{\mu\nu} =
\partial^2_{\alpha\beta}H^{\mu\alpha\nu\beta},~~H^{\mu\alpha\nu\beta} =
H^{\nu\beta\mu\alpha} = -H^{\beta\nu\mu\alpha}
\label{16}
\eeq
since $\Delta^{\mu\nu}$ is identically conserved, symmetric and does not
alter the
Lorentz generators for arbitrary (aymptotically well-behaved) functions
$H$ with the
above symmetries.  This
is equivalent to adding non-minimal gravitational couplings
$\sim R_{\mu\alpha\nu\beta}H^{\mu\alpha\nu\beta}, R_{\mu\nu}H^{\mu\nu}$
or $RH$ to the original minimal action.  Thus, our task is to find
$\Delta^{\mu\nu}$ whose traces cancel those of the original
$T^{\mu\nu}$, at least
on shell.  If this can be accomplished  in a given gauge, the theory
will have
conserved dilatation and conformal currents (\ref{3}), (\ref{5})
[This is the basis of the scalar theory's improvement, using
$\Delta_{\mu\nu} =
\frac{1}{4}(d-2)(d-1)^{-1}( \eta_{\mu\nu}\Box - \partial^2_{\mu\nu})
\phi^2.]$

We begin with vector theory.   Since the Maxwell tensor's trace is
$(1-d/4)F^2_{\mu\nu}$, $H$ would have to be bilinear in the vector
potentials.  In option a), we use the fact that as for any
free Abelian gauge theory, we can express all
quantities entirely in terms of the physical modes;  here
the transverse spatial components $({\bf A}^T,\dot {\bf A}^T), {\bf
\nabla} \cdot
{\bf A}^T \equiv 0$, due to the Gauss constraint.
Dropping the ``$T$" notation henceforth, we write
  \bea T^\alpha_\alpha &=&
(1-\frac{d}{4})F^2_{\mu\nu} = \frac{1}{2}(d-4)
\left[ {\bf \dot A}^2 - (A_{i,j})^2 + \partial^2_{ij}(A_iA_j) \right]
\nonumber\\ &
\cong & \frac{1}{4}(d-4)\left[ 2\partial^2_{\mu\nu} - \eta_{\mu\nu}\Box
\right](A^\mu
A^\nu)~,~ A_\mu \equiv (0,A_i)~.
\label{17}
\eea
The last equality is valid on shell $(\Box{\bf A}=0)$ and the space-time
notation is purely for convenience.  It is easy to check that if we
introduce
\bea
H^{\mu\alpha\nu\beta} &=& a \eta^{\mu\nu}A^\alpha A^\beta
+ b
\eta^{\mu\nu}\eta^{\alpha\beta}A^2 + {\rm symm.} \nonumber \\
a &=& \frac{(4-d)}{2(d-2)}~,\quad\quad\quad b = \frac{d(d-4)}
{4(d-1)(d-2)}~.
\label{18}
\eea
the corresponding $\Delta^\mu_\mu$ just cancels (\ref{17}).  The
dilatation generator $D$ is likewise improved, to the desired form, $D =
-\int
\dot{\bf A} \left(\frac{d-2}{2} + r \cdot \nabla \right) {\bf A}
d^{d-1}r$, as is
$K^0$, which acquires a non-moment term $= \frac{d-2}{2} \int {\bf A}^2
d^{d-1}r$;
finally, $K^i$ is ``improved" in that the part proportional to $-\int
x^i\dot
{\bf A} \cdot {\bf A}$ acquires an overall $(d-2)$ coefficient, exactly
like
the scalar field's improvement.  [The special case $d=2$ of course
corresponds to
the trivial theory in which there is no ${\bf A}^T$ at
all!]  However, the various $\Delta^{\mu\nu}$ components are not Lorentz
covariant and the improved generators have the ``wrong" commutators with
the
(unaltered) Lorentz generators.  This can be seen upon
reintroducing the usual
longitudinal and temporal components of $A_\mu$ through setting $A_\mu
\rightarrow A_\mu + \partial_\mu \alpha$, thereby shifting the values of
$\Delta^0_\mu$.  This difficulty incidentally shows the limitation on
the
folklore that a free Abelian gauge theory can be entirely treated in
terms of its
transverse, physical, modes with Lorentz invariance always taking care
of itself.

If we retain manifest Lorentz invariance, as in option b),
 keeping the usual
$A_\mu$, then
 \bea
 T^\alpha_\alpha &=& 2\left(1-\frac{d}{4}\right)
\partial_\mu       A_\nu(\partial^\mu A^\nu - \partial^\nu A^\mu)
\nonumber \\
&\cong& 2\left(1-\frac{d}{4}\right)\left[ \left( \eta_{\mu\nu}\Box - 2
\partial^2_{\mu\nu}
\right) A^\mu A^\nu + 2 \partial_\mu(A^\mu \partial_\nu A^\nu)\right]
\label{19}
\eea
The last term in (\ref{19}) is a non-superpotential residue that cannot
be
cancelled without gauge-fixing, in accordance  with our general
arguments.
In Landau gauge the additional term vanishes, and $\Delta^{\mu\nu}$
can be chosen as in Eqs.~(\ref{17}), (\ref{18}). In Feynman gauge the
improved
 tensor will be simply given by d copies of
 the tensor for scalars, where the Lorentz index of the vector field
 acts like an internal symmetry index.
If one adds the required FP ghosts one has a conformally
 invariant action,
formally
equivalent to the original one. The price here is, of course, that
conformal inversion does not commute with the BRS operator. Therefore
the symmetry
does not act within the physical Hilbert space and so this improved
tensor is not
an allowed operator.

There is a simple way, related to our group theoretical arguments, to
see why
scale invariance cannot be reinstated:  the scale dimension of $A_\mu$,
as
obtained by acting on it with $D$, must be compatible with gauging
$A_\mu
\rightarrow A_\mu + \partial_\mu\alpha$;  but $\alpha$, being
non-dynamical, has
scale dimension zero.  Hence $A_\mu$ must have dimension unity, leading
to $d_c =
4$.  The same problem holds for the antisymmetric tensor field away from
$d=6$.
The simple argument here is that the potential $B_{\mu\nu}$ is invariant
under
$B_{\mu\nu} \rightarrow B_{\mu\nu} + (\partial_\mu\lambda_\nu -
\partial_\nu\lambda_\mu)$, i.e., under a transverse vector, $\lambda^\mu
\epsilon^{\mu\nu}...\partial_\nu\Lambda ...$, so that really $\delta
B_{\mu\nu}
\sim \partial^2\Lambda$ and it must have dimension 2, compatible only
with $d_c =
6$.

For the antisymmetric tensor theory, whose Lagrangian is
$-1/6~H^2_{\mu\nu\alpha}$,
the stress tensor trace is
\beq
T^\alpha_\alpha = (1 -
\frac{d}{6})H^2_{\mu\nu\alpha}\quad\quad\quad H_{\mu\nu\alpha} =
\partial_{[\alpha}B_{\mu\nu]}~.
\label{20}
\eeq
the covariant from corresponding to (\ref{19}) is
\bea
T^\alpha_\alpha &=& \frac{1}{2} (6-d) \partial_\alpha \{
B_{\mu\nu}(\partial_\alpha
B_{\mu\nu} + 2\partial_\nu B_{\alpha\mu} \} \nonumber \\
&\cong & (6-d) \left[ \frac{1}{4} \Box B^2_{\mu\nu} +
\partial^2_{\nu\alpha}
(B_{\mu\nu} B_{\alpha\mu}) - \partial_\alpha (B_{\alpha\mu} \partial_\nu
B_{\mu\nu}) \right]
\label{21}
\eea
with a single derivative term vanishng only in Lorentz gauge
$\partial_\nu B_{\mu\nu} = 0$.  In terms of the physical mode,
$T^\alpha_\alpha$ \underline{is} a double derivative, however.  Here,
the gauge invariant (but not Lorentz scalar) field is $\varphi$, where
$B^{ij} =
\epsilon^{ijk}\partial_k \varphi$, working in $d=4$ for simplicity.
It follows by use of the constraint $\partial_kH^{0ik} = 0$ that
$T^\alpha_\alpha$ is simply proportional to $\Box \dot\varphi^2$ (or to
$\Box(\partial_\mu \varphi)^2$ plus a superpotential) so that one may
use
\beq
\Delta_{\mu\nu} \sim (\eta_{\mu\nu} \Box - \partial^2_{\mu\nu}) \dot
\varphi^2
\label{22}
\eeq
to cancel $T^\alpha_\alpha$.  [This discussion and the final form
(\ref{22}) are actually valid for all dual models.]  Covariantizing
(\ref{22})
by  reinstating the gauge parts of $B_{\mu\nu}$, however, again reveals
its gauge
variance;  one cannot have both symmetries simultaneously away from
$d=6$. We
shall return to this point below.

\section{Dualities}
It is well known, and leads to a well-known paradox (see for example
\cite{bb}),
that in $d=4$ the $N=2$ action is equivalent to that of a massless
scalar, as is
the Maxwell action in $d=3$ and more generally the $N$-form's action in
$d = N+2$
where the dual field strength is a 1-form $^{\ast}H$.  For, consider
there the
Lagrangian  \beq
{\cal L} = - \frac{1}{2} A^2_\mu + {^*H}^\mu A_\mu~.
\label{23}
\eeq
Eliminating $A_\mu$ gives ${\cal L} \sim {^\ast}H^2$, while varying with
respect to the (Lagrange multiplier) $B_{\alpha\beta}$ field
tells us that $A_\mu$ is the gradient of a
scalar $\Psi$, and ${\cal L} \sim (\partial_\mu \Psi)^2$.  The
gauge-invariant
stress tensor is
\beq
T^{\mu\nu} \sim {^\ast}H^{\mu~\ast}H^{\nu} - \frac{1}{2}
g^{\mu\nu}(^{\ast}H)^2~,~~T^\alpha_\alpha \sim - ^{\ast}H^2~.
\label{24}
\eeq
It is conserved by virtue of the Bianchi identity
$\partial_\mu~{^\ast}H^\mu =
0$.  The only way to improve it without losing gauge or Lorentz
invariance would
seem to be a non-local one.  Indeed, we may postulate
\beq
\Delta_{\mu\nu} = \frac{1}{3} (\eta_{\mu\nu} \Box - \partial^2_{\mu\nu})
\Box^{-1}(^\ast H)^2~.
\label{25}
\eeq
This apparently  unacceptable non-local operator has a saving feature,
however:
its matrix elements between zero mass gauge invariant states are
\underline{local}, the $\Box^{-1}$ is cancelled.  Since the only gauge
invariant
operators in the theory are $H^*$ and expressions built from them, the
simplest way to verify this is to consider the Fourier transform
of the correlator $C_{\mu\nu\alpha\beta}$ of $\Delta_{\mu\nu}(x) $ with
$H^*_\alpha(y)$ and $H^*_\beta(z)$. $\Delta$ carries momentum $q$ and
the
$H^*$'s  $k_1$
 and $k_2$ , respectively. Taking the discontinuity in $k_1$ and
 $k_2$ puts the states on mass shell and we obtain:
 \beq
 disc C_{\mu\nu\alpha\beta} \sim \delta(k^2_1) \delta(k^2_2) k_{1\alpha}
k_{2\beta} (\eta_{\mu\nu} q^2-q_\mu q_\nu)
 \label{26}
\eeq
using the gauge invariant propagator
$\langle{^\ast}H^\alpha~{^\ast}H^{\beta}\rangle \sim k^{-2}(k^\alpha
k^\beta -
g^{\alpha\beta}k^2)$.  To go off mass shell, we simply complete
(\ref{26}) to
 its full analytic structure, i.e. the correlator is given by:
\beq
C_{\mu\nu\alpha\beta} \sim (k^2_1k^2_2)^{-1} k_{1\alpha}k_{2\beta}
 (\eta_{\mu\nu}q^2-q_\mu q_\nu)~.
\label{27}
\eeq
Naive use of the gauge invariant propagator would add terms
singular in $q^2$ and proportional  to $k^2_1$ or $k^2_2$. Discarding
these terms is completely consistent with analyticity, gauge invariance
and Lorentz invariance, but not with the Bianchi identity off-shell,
and it corresponds to the (required) use of the equation of motion
$(k^2_i = 0)$ in
the scalar improvement.

Once the expression for the basic correlator (\ref{27}) is given,
calculating
that of any number of stress tensors is straightforward, and the result
is
identical to that obtained in terms of the improved scalar:  we are
simply realizing
the $^\ast H^\alpha \sim \partial^\alpha \Psi$ correspondence in the
Feynman
diagrams.  In particular, the trace anomalies
                        calculated with the
improved tensor
in the scalar or $N$-form language automatically have the same values.
Actually,
since the anomaly depends only on mass-shell matrix elements \cite{cc},
one can
prove the equality starting from (\ref{26}).

 The improvement
presented here has its counterpart in operator language, where we saw at
Eq.
(\ref{22}) how in the physical sector (obeying the Gauss constraint) of
Hilbert
space, the improvement becomes local.  We will discuss elsewhere the
formulation of
improvement in an external gravitational field and the realization of
the Weyl
group.  We also remark that a similar treatment is possible in the
trivial case of
$d = N+1$ and the $N$-form is ``dual to nothing".  Here our procedure
would simply
yield a vanishing correlator of the improved stress tensor with any
gauge invariant
operator.  Finally, we mention that in addition to the usual massless
scalar
action, there are two others that are scale or Weyl invariant at $d=4$,
although
they are less physical.  Both have $N = 2$ counterparts.  We simply list
them
here.  The first is ${\cal L} = -\frac{1}{2} \Psi\Box^2\Psi$, which can
be locally
improved, but represents a dipole ghost excitation.  Its counterpart is
non-local.
The other is ${\cal L} = [(\partial_\mu \Psi)^2]^2$; its $N=2$ partner
is
a non-polynomial action ${\cal L} \sim (^{\ast}H^2)^{2/3}$.  Both
actions are
manifestly  Weyl invariant, being homogeneous of degree zero in the
metric.

\section{Conclusion}
We have seen, by general arguments and by specific examples that
improving $N$-form
actions away from critical dimension required either non-locality  or
loss of
Lorentz or gauge invariance.  However, for dimension $d = N+2$ where the
$N$-form is dual to a scalar, we saw that improvement can be carried out
quasilocally (on-shell).  This possibility provided the resolution of
the
ancient ``paradox" of conformal non-invariance on the $N$-form side, and
in
particular explained the equality of the conformal anomalies of both
partners.

\vspace{1.0cm}

This work was supported by NSF grants PHY-93-15811, 92-45316, by
Israel-US
Binational grant 89-00140, and by the Ambrose Monell  Foundation.

  \end{document}